\documentclass[prl,twocolumn,superscriptaddress,floatfix]{revtex4}
\usepackage{comment}
\usepackage{hyperref}
\usepackage[all]{hypcap}
\usepackage{lipsum}
\usepackage[british]{babel}
\usepackage[utf8]{inputenc}
\usepackage{amsmath}
\usepackage{amssymb}
\usepackage{mathtools}
\usepackage{braket}
\usepackage{graphicx}
\usepackage[dvipsnames]{xcolor}
\usepackage{braket}
\usepackage{appendix}
\usepackage[normalem]{ulem}
\makeatletter
\newcommand*\bigcdot{\mathpalette\bigcdot@{.5}}
\newcommand*\bigcdot@[2]{\mathbin{\vcenter{\hbox{\scalebox{#2}{$\m@th#1\bullet$}}}}}
\makeatother
\makeatletter
\renewcommand{\thetable}{\@arabic\c@table}
\makeatother

\begin{document}
\title{Non-Relativistic Hyperboloidal Method for Quasinormal Modes}
\title{Hyperboloidal Method for Quasinormal Modes of Non-Relativistic Operators}
\author{Christopher Burgess}
\author{Friedrich K\"{o}nig}
\email{fewk@st-andrews.ac.uk}
\affiliation{School of Physics and Astronomy, SUPA, University of St. Andrews,
North Haugh, St. Andrews, KY16 9SS, UK}
\date{\today}

\begin{abstract}
The recently reported compactified hyperboloidal method has found wide use in the numerical computation of quasinormal modes, with implications for fields as diverse as gravitational physics and optics. We extend this intrinsically relativistic method into the non-relativistic domain, demonstrating its use to calculate the quasinormal modes of the Schr\"{o}dinger equation and solve related bound-state problems. We also describe how to further generalize this method, offering a perspective on the importance of non-relativistic quasinormal modes for the programme of black hole spectroscopy.
\end{abstract}

\maketitle

\vspace{-0.2cm}
\section{Introduction}
\vspace{-0.2cm}
Quasinormal modes (QNMs) are complex frequency modes which characterize the resonant response of a system to linear perturbations.
They are prevalent in the physics of waves, with special prominence in optics and gravitational physics.
In optics, QNMs are useful for understanding the behaviour of resonant photonic structures, such as plasmonic crystals, nanoparticle traps, metal gratings, and optical sensors \cite{Lalanne2018, Yan2018, Qi2021, Gras2019, Ren2021}.
In gravitational physics, they are thought relevant to tests of black hole no-hair conjectures \cite{Gossan2012, Shi2019, Ma2023}, and central to the emerging project of black hole spectroscopy with gravitational waves \cite{Dreyer2004, Cabero2020}.
While the QNM literature in optics treats dispersion as a matter of necessity \cite{lalanne2019, Primo2020}, the prevailing methods in gravitational physics are concerned with non-dispersive, relativistic wave propagation \cite{Kokkotas1999, Berti2009, Konoplya2011}.
We believe there are good reasons to go beyond relativistic wave propagation in the gravitational context.
A variety of quantum gravity models predict the dispersive propagation of gravitational waves \cite{Nishizawa2018, Mastrogiovanni2020, Ezquiagal2021, Aydogdu2022}, for example, in models with a non-zero graviton mass, violation of Lorentz invariance, and higher dimensions \cite{Will1998, Sefiedgar2011, Kostelecky2016}.
Indeed, it has been proposed that QNMs may be used to probe gravity beyond general relativity, through imprints on radiative emission from black holes \cite{Glampedakis2019, Agullo2021, Srivastava2021, Chen2021, Fu2024}.
More generally, we anticipate that development of QNM methods for non-relativistic operators will broaden the scope of existing questions in QNM theory.

Numerical methods underpin much of the progress in QNMs over recent years.
Indeed, efficient schemes for computing the QNMs of potentials are likely indispensable for future developments in both theory and the modelling of observations.
Recently, the so-called compactified hyperboloidal method \cite{Zenginoglu2008, Zenginoglu2011, Jaramillo2021, Macedo2024} has proven to be a powerful tool, finding wide use in the computation of black hole QNM spectra and bringing within reach the systematic exploration of their connection to pseudospectra \cite{Destounis2021, Ripley2022, Gasperin2022, Sarkar2023, Boyanov2024, Cao2024, Zhu2024, Destounis2024}.
Beyond this, it is natural to ask whether the method can also find use in optical systems.
We believe it can, but it cannot be widely applied in optics without modification.
This is because optical media create non-relativistic and dispersive dynamics, while the present formulation of the method treats only relativistic and non-dispersive dynamics, as may be seen from its use of hyperbolic spatial slices penetrating the black hole horizon and future null infinity.

A notable optical system that motivates the development of a hyperboloidal method for optics is the fiber optical soliton, which has recently been established as a black hole analogue with an exactly known QNM spectrum \cite{Burgess2024}.
As such, the soliton is the ideal system with which to develop the method, as the resulting numerics can be compared both to known analytical results and to the numerics of the corresponding relativistic system.
Moreover, perturbations to the soliton realize the Schr\"{o}dinger equation with a P\"{o}schl-Teller potential, making the soliton a promising experimental platform with which to address questions in QNM theory, such as the physical status of spectral instabilities observed in QNM numerics, where the P\"{o}schl-Teller potential is paradigmatic \cite{Ferrari1984a, Nollert1993, Cho2012, Jaramillo2021, Cardoso2024}.

In this article, we outline a new method for the numerical computation of QNM spectra for operators with a non-relativistic dispersion relation, by adapting the compactified hyperboloidal method.
We begin by showing how to compute the QNMs of the Schr\"{o}dinger equation for an arbitrary potential, noting that the relativistic and non-relativistic spectra are related by a simple endomorphism. We subsequently demonstrate the method for the P\"{o}schl-Teller potential, explicitly calculating the soliton QNM spectrum numerically.
Finally, we sketch how to develop these ideas in order to treat generalized non-relativistic dispersion relations, and discuss potential applications of the more general method, with emphasis on its future use in black hole spectroscopy.

\vspace{-0.2cm}
\section{Compactified Hyperboloidal Method for the Schr\"{o}dinger Equation}
\vspace{-0.2cm}
We begin by considering a scalar field $\phi$ which obeys a Schr\"{o}dinger equation of the form,
\begin{equation}
(i\partial_t - \partial_r^2 + V)\phi = 0,\label{schrEqn}
\end{equation}
with $V$ a potential that vanishes for $r\to\pm\infty$.
The boundary conditions for QNMs describe solutions that transport energy away from the potential, as discussed in more detail in \cite{Burgess2024}.
It can be shown, using the asymptotic dispersion relation of Eq.~(\ref{schrEqn}), that QNM solutions must diverge for $r\to\pm\infty$.
That is, the asymptotic form of the solution must be $\phi \sim \exp(iKr - i\Omega t)$, with the requirement that $\text{Im}(K)$ is positive and negative on the left and right, respectively.
These spatial divergences are problematic for numerical methods, but they can be removed by using a hyperboloidal coordinate transformation.
Following \cite{Jaramillo2021}, we adopt coordinates,
\begin{equation}
t = \tau - h(x), \quad r = g(x)\label{coordtransformation},
\end{equation}
where $g(x), h(x)$ are yet to be given, and $\partial_\tau = \partial_t$ by construction.
In the relativistic context, these are used to compute QNMs of black holes, with $h(x)$ chosen so that contours of $\tau$ tend to null curves that intersect the horizon and future null infinity.
There, Eq.~(\ref{coordtransformation}) is intended to respect the asymptotic hyperbolic geometry of the spacetime, giving rise to bounded and well-behaved QNM solutions.
However, there is no preferred speed in our non-relativistic system, meaning that no coordinate transformation will consistently give rise to bounded solutions. This requires a different approach.

In order to construct bounded QNM solutions, we first parameterize $h(x)$ by a new variable $v_g$ such that contours of $\tau$ tend asymptotically to trajectories directed outwards with $|dr/dt| = v_g$. In particular, we write
\begin{equation}
g(x) = \text{arctanh}\ \!{x}, \quad h(x) = (2v_g)^{-1} \log(1-x^2)\label{coordfuncs},
\end{equation}
where $g(x)$ compactifies the space such that the real line of $r$ gives $x \in [-1,1]$ if we close the set by including the boundaries.
In the Appendix, we show that QNMs whose asymptotic group velocity is $v_g$ in $(r, t)$ coordinates are finite at the spatial boundaries, $x = \pm 1$.
This enforces the boundary conditions for these modes, but does not guarantee that any such modes exist.

In contrast to the relativistic case, dispersion in non-relativistic systems means that group and phase velocities are not the same.
As a result, QNMs whose asymptotic phase velocity is $v_p \neq v_g$ in $(r, t)$ coordinates will undergo phase divergences at the boundaries.
This can be removed by a phase-rotation of the field,
\begin{equation}
\hat{\phi} = e^{-\Delta\log(1-x^2)/2}\phi\label{phaserotation},
\end{equation}
where we introduce $\Delta = i(v_g - v_p)/2$, so that the phase rotation is parameterized by both $v_g$ and $v_p$. The form of the required phase rotation follows from the asymptotic dispersion relation of Eq.~(\ref{schrEqn}) and the choice of height function, $h(x)$. Intuitively, it depends on the mismatch of the two velocities.
The result is that the field $\hat{\phi}$ is bounded and well-defined on the new space.

The cost of the above construction is that we introduce two unknown real parameters, $v_g$ and $v_p$, into the problem.
In fact, identifying velocity pairs that correspond to actual QNM solutions is as difficult a problem as determining the QNM spectrum itself. This may be seen by the relation,
\begin{equation}
\Omega = -\frac{1}{2}v_g\bigg(v_p + i\sqrt{v_g(v_g-2v_p)}\bigg),\label{FrequencyInTermsOfSpeeds}
\end{equation}
which we derive, in the Appendix, from the asymptotic dispersion relation of Eq.~(\ref{schrEqn}).
This holds true for any mode whose asymptotic group and phase velocities are $v_g$ and $v_p$, respectively.
The existence of a relation such as Eq.~(\ref{FrequencyInTermsOfSpeeds}) is a direct consequence of dispersion.
In a relativistic system, all asymptotic speeds are the speed of light, so $\Omega$ cannot be expressed in terms of asymptotic velocities.
This difference between the relativistic and non-relativistic methods is crucial.
Eqs.~(\ref{coordfuncs}) and (\ref{phaserotation}) mean we obtain an equation of motion, and an eigenvalue equation for the complex frequency $\Omega$, both of which are parameterized by $v_g$ and $v_p$.
The significance of Eq.~(\ref{FrequencyInTermsOfSpeeds}) is that these additional parameters can ultimately be eliminated, leaving $\Omega$ as the only unknown in the problem.

We proceed as in \cite{Jaramillo2021}, by rewriting Eq.~(\ref{schrEqn}) in the new coordinates and performing a first-order reduction in time, introducing the auxiliary field $\hat{\psi}$.
The equation of motion becomes $\partial_\tau\hat{\phi} = \hat{\psi}$ with
\begin{equation}
x^2\partial_\tau\hat{\psi} = J_1\hat{\phi} + J_2\hat{\psi},\label{InitialEquationOfMotion} 
\end{equation}
where $J_1$ and $J_2$, given in the Appendix, are spatial operators depending on the potential and the asymptotic velocities.
In contrast to the relativistic method, $\partial_\tau\hat{\psi}$ cannot be isolated by division in Eq.~(\ref{InitialEquationOfMotion}) because its pre-factor vanishes at $x = 0$.
This occurs because the contours of $\tau$ have to ``turn around'' in order to be outgoing in the $(r, t)$ coordinates.
Our alternative approach is to construct $\partial_\tau\hat{\psi}$ using a Taylor series around $x = 0$, obtaining the required derivatives by repeated differentiation of Eq.~(\ref{InitialEquationOfMotion}).
In fact, this treatment is necessary only for terms indivisible by $x^2$, and we obtain a simpler result if we initially separate the terms in this way.
This separation is mostly trivial, but for the potential, where we write $V = V_0 + xV_1 + x^2\tilde{V}(x)$, with $V_0$ and $V_1$ Taylor series coefficients about $x = 0$ and $\tilde{V}$ accounting for the remaining terms.
We obtain
\begin{equation}
\partial_\tau\hat{\psi} = L_1\hat{\phi} + L_2\hat{\psi},\label{EquationOfMotion}
\end{equation} 
where $L_1$ and $L_2$ are spatial operators that we derive in the Appendix.
Eq.(\ref{EquationOfMotion}) is formally identical to that obtained in the relativistic method \cite{Jaramillo2021}, but the operators are quite different, containing arbitrarily high spatial derivatives and depending on the asymptotic velocities.

In matrix form, we write $i\partial_\tau u = L u$ with
\begin{equation}
u \equiv
\begin{pmatrix}
\hat{\phi} \\
\hat{\psi}
\end{pmatrix},\quad
L \equiv
i\begin{pmatrix}
0 & 1\\
L_1 & L_2
\end{pmatrix},\nonumber
\end{equation}
and obtain the mode equation
\begin{equation}
L u = \Omega u.\label{EigenvalueProblem}
\end{equation}
The operator $L$ is parameterized by $v_g$ and $v_p$, giving rise a family of operators.
For each operator, Eq.~(\ref{EigenvalueProblem}) defines a unique eigenvalue problem and a corresponding spectrum.
However, only a subset of the frequencies from these spectra obey Eq.~(\ref{FrequencyInTermsOfSpeeds}), and it is this subset which comprises the QNM spectrum of Eq.~(\ref{schrEqn}).
Using Eq.~(\ref{FrequencyInTermsOfSpeeds}) to eliminate $v_g$ and $v_p$, one obtains a problem in which the frequency $\Omega$ is the only unknown and all solutions correspond to QNMs.
In this formulation, $L$ is parameterized by $\Omega$, which constitutes an essential difference from the relativistic method, wherein the corresponding operator does not depend on $\Omega$ \cite{Jaramillo2021}.
Importantly, Eq.(\ref{EigenvalueProblem}) unambiguously determines the QNM spectrum.

Eq~(\ref{EigenvalueProblem}) is discretized using $N$-point Chebyshev nodes of the second kind.
In this way, fields are approximated by $N$-dimensional vectors and spatial operators by $N$-dimensional matrices. It follows that the vector $u$ and the operator $L$ are approximated by $2N$-dimensional vectors and matrices, respectively. 
The result is
\begin{equation}
L^{N} u^N = \Omega \ \! u^N.\label{DiscretizedEigenvalueProblem}
\end{equation}
The QNM spectrum may then be obtained from Eq.~(\ref{DiscretizedEigenvalueProblem}) in the usual way using $\det(L^N - \Omega\ \!\text{Id}) = 0$.
In the Appendix, we show that this determinant may be rewritten as that of a smaller $N$-dimensional matrix, $M$.
Its elements are quadratic in the square root of the QNM frequency, giving rise to a polynomial of degree $2N$ in $\sqrt{\Omega}$.
For a given potential $V$, the roots may be numerically determined in order to give $2N$ of the QNM frequencies.
The fact that the frequency enters via its square root is a result of the Schr\"{o}dinger equation having a first derivative in time, rather than a second derivative in time like the wave equation.
Indeed, the exact QNM spectra of the Schr\"{o}dinger and wave equations are related to each other by $i\sqrt{\Omega} = \omega$, as was elaborated in \cite{Burgess2024}.
This means we can relate the results of the non-relativistic method to those of the relativistic method, allowing us to better evaluate the accuracy of the new method.

\vspace{-0.2cm}
\section{Quasinormal Modes of the P\"{o}schl-Teller Potential}
\vspace{-0.2cm}
In this section, we use the above numerical method to calculate the QNMs of the Schr\"{o}dinger equation with the P\"{o}schl-Teller potential,
\begin{equation}
V = V_0\ \!\text{sech}^2(r) = V_0(1 - x^2),\label{UnperturbedOrPerturbedPotential}
\end{equation}
which serves as an exemplar for both the relativistic and non-relativistic methods.
The QNMs of Eq.~(\ref{UnperturbedOrPerturbedPotential}) are finite polynomials in the compactified spatial coordinate, with the result that an $N$-point discretization reproduces the first $2N$ QNMs to arbitrary precision.
The P\"{o}schl-Teller potential is also ideal because the corresponding QNM spectrum of the Schr\"{o}dinger equation is given analytically by
\begin{align}
\Omega_n = \Bigg[\!\ n+\frac{1}{2} - i\sqrt{V_0-\frac{1}{4}}\!\ \Bigg]^{\!2}\!\!,\label{QNMFrequencies}
\end{align}
allowing us to verify our results \cite{Burgess2024}.
In regards to the non-relativistic method, we note that the P\"{o}schl-Teller potential is especially simple because all its QNMs have the same $v_g$ parameter, which is a result of the fact that $i\sqrt{\Omega}$ is aligned along vertical lines in the complex plane for this potential.
While this simplicity does not influence the operation of the method, it does allow us to more easily assess the spectrum.
Lastly, we partition the P\"{o}schl-Teller potential with $V_1 = 0$ and $\tilde{V} = -V_0$, which reinforces the simplicity of the potential.

\begin{figure}
\centering
\includegraphics[width=245pt]{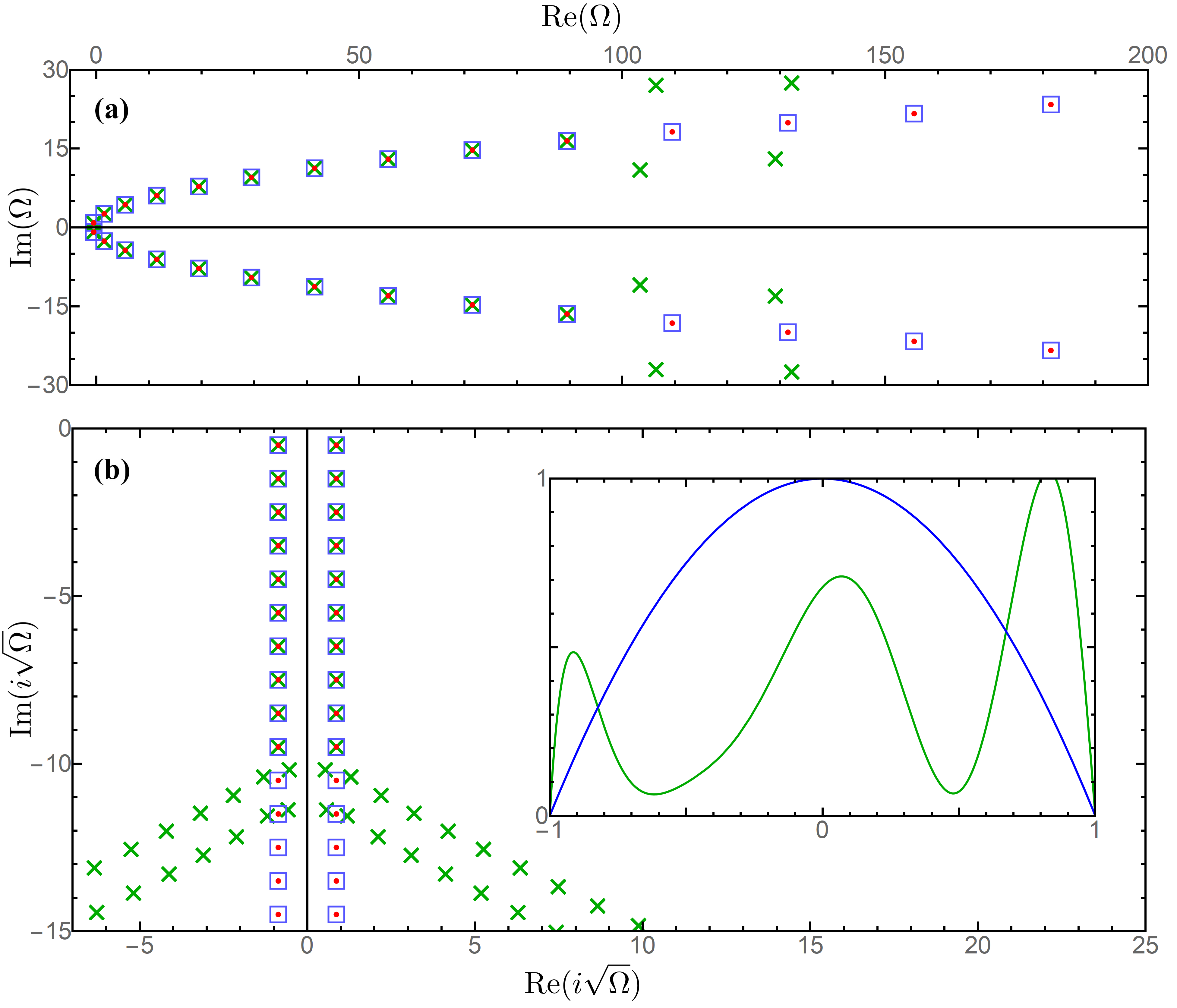}
\caption{
QNM spectra for the Schr\"{o}dinger equation with a potential $V = V_0\hspace{1.5pt}\text{sech}^2(r) + \epsilon\Delta V$, where $V_0 = 1$ and $\Delta V$ is the perturbation shown in green in the inset, alongside the unperturbed P\"{o}schl-Teller potential in blue. The red dots and blue boxes correspond to the unperturbed P\"{o}schl-Teller potential ($\epsilon = 0$), with red dots ({\color{red}•}) given by the exact formula and blue boxes ({\color{blue}$\square$}) numerically determined by the new method. The green crosses ({\color{ForestGreen}$\times$}) correspond to a perturbed potential ($\epsilon = 10^{-30}$) and are also numerically determined. \textbf{(a)} displays the three QNM spectra, while \textbf{(b)} displays the same spectra under the transformation $\Omega \to i\sqrt{\Omega}$, which relates the spectra to those of a corresponding relativistic operator.}
\label{fig:1}
\end{figure}

Now, we make some comments on the specifics of our implementation of the method.
We find the calculation is significantly more efficient for odd $N$.
This is a consequence of discretization.
The Taylor series expansions of $L_1$ and $L_2$ involve spatial derivatives at $x = 0$, which are obtained by integration with a Dirac delta function in the continuous case, and by matrix multiplications in the discretized case.
For odd $N$, the relevant matrix is zero everywhere but a central column whose entries are unity.
However, for even $N$, the matrix is everywhere populated, and this increases the computational cost of the calculation.
We also find that evaluating the determinant of the large symbolic matrix $M$ is inefficient, so we instead sample the determinant in the complex plane and reconstruct the symbolic determinant using polynomial interpolation.
This uses that the method produces a polynomial of degree $2N$ in $\sqrt{\Omega}$.
Importantly, this is true no matter what potential we consider.

\begin{figure}
\centering
\includegraphics[width=245pt]{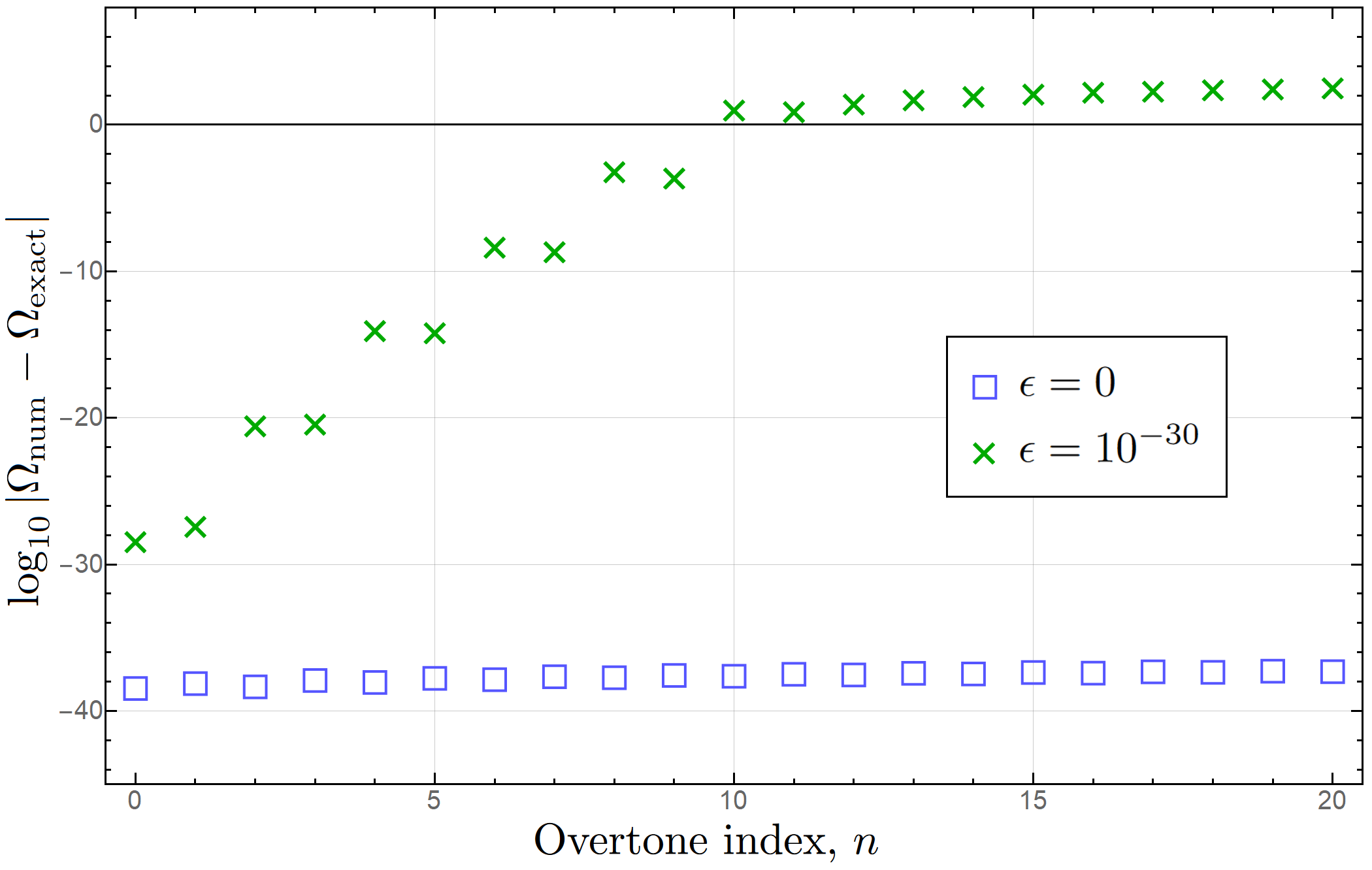}
\caption{
Comparisons of exact and numerically determined QNM frequencies for the Schr\"{o}dinger equation with a potential $V = V_0\hspace{1.5pt}\text{sech}^2(r) + \epsilon\Delta V$, where $V_0 = 1$ and $\Delta V$ is the perturbation shown in Figure~1. The first 21 QNM frequencies are displayed. The unperturbed ($\epsilon=0$) spectrum is recovered well by the new numerical method, with errors smaller than $10^{-37}$ for the chosen working precision. We also obtain the perturbed ($\epsilon=10^{-30}$) spectrum and find the deviation from the exact spectrum grows rapidly with overtone index, $n$, as in previous works on spectral instability.}
\label{fig:2}
\end{figure}

In Figure~\hyperref[fig:1]{1(a)}, we plot the exact QNM frequencies of the unperturbed P\"{o}schl-Teller potential, given in Eq.~(\ref{QNMFrequencies}) \cite{Burgess2024} alongside those calculated by the new numerical method, with a resolution of $N = 201$.
We find excellent agreement for all frequencies, with an error which may be made arbitrarily small by increasing the working precision.
These results are given in Figure~\hyperref[fig:2]{2}.
We also calculate the QNM spectrum of a perturbed P\"{o}schl-Teller potential,
\begin{equation}
V = V_0(1-x^2) + \epsilon\Delta V,\label{PerturbedPotential}
\end{equation}
where $\epsilon = 10^{-30}$ and $\Delta V$ is a randomly chosen polynomial of degree 9, shown in the inset of Figure~\hyperref[fig:1]{1}.
We find that the spectrum for Eq.(\ref{PerturbedPotential}) closely resembles the unperturbed spectrum up to the $10^{\text{th}}$ overtone index, beyond which the frequencies are significantly displaced from their unperturbed values, as shown in Figure~\hyperref[fig:2]{2}.
These numerical results are then indicative of spectral instabilities that have been reported by previous authors \cite{Nollert1996, Daghigh2020, Jaramillo2021}.

The simple relationship between the QNMs of the Schr\"{o}dinger and wave equations becomes visible under the transformation $\Omega \to i\sqrt{\Omega}$, which maps the spectrum of the former onto that of the latter.
In Figure~\hyperref[fig:1]{1(b)}, we plot $i\sqrt{\Omega}$ for the same spectra as above, obtaining the recognizable vertical lines in the complex plane that are characteristic of the wave equation with a P\"{o}schl-Teller potential. In this way, we illustrate how one can cross-verify the results of the relativistic and non-relativistic methods against each other, for arbitrary potentials.

\vspace{-0.2cm}
\section{Discussion}
\vspace{-0.2cm}
In this section, we discuss potential applications of the non-relativistic compactified hyperboloidal method that we developed in the preceding text, suggesting well-motivated directions in which to further develop the method and providing a sketch of how this can be achieved.
The main motivations for this method were the modelling of QNMs of optical solitons, and the development of a framework within which one can treat QNMs in quantum gravity models with dispersive gravitational wave propagation.
Beyond these, we note that this non-relativistic method may be employed equally well in any system governed by a Schr\"{o}dinger equation equipped with a general potential.
In this paper, we numerically calculated QNM spectra for the P\"{o}schl-Teller potential and perturbations of that potential, finding agreement with earlier works \cite{Boonserm2011, Cardona2017, Burgess2024}.
For potentials with different long-range behaviour than the P\"{o}schl-Teller potential, one typically requires different choices of height function $h(x)$, but this requirement is shared by the relativistic method, and may be addressed by the same techniques \cite{Zenginoglu2011, Jasiulek2011, Macedo2018}. 
In addition, we note that this method may also be used to numerically solve for the quantum mechanical bound-states of a general potential well, using the well-known connection between the QNMs of a potential barrier and the bound-states of the corresponding well \cite{Ferrari1984a, Ferrari1984b, Churilova2022, Volkel2022}.

As described above, the non-relativistic method we have presented is closely related to the relativistic method, sharing many essential features with it.
For instance, the classes of potentials that can be treated by the two methods are the same, and they have the same maximum achievable accuracy for a given resolution.
As a result, the methods are comparable in their scope and power.
They also share the same advantages and disadvantages when compared to other popular numerical methods, such as Leaver’s continued fraction method \cite{Leaver1986}. For example, in this case, both the relativistic and non-relativistic methods enjoy the advantage that they recover the entire spectrum simultaneously, and do not require initial seed values close to the QNM frequencies one wishes to compute \cite{Leaver1986, Zenginoglu2009, Ansorg2016, Jaramillo2021}.

The non-relativistic method we have presented readily generalizes beyond the Schr\"{o}dinger equation, allowing us to treat a large class of more general non-relativistic operators.
Indeed, the method presented in this paper primarily serves a didactic purpose, as a demonstration of a general approach with which one may calculate QNMs of these more general operators.
The primary motivation for this is to facilitate the efficient computation of QNMs of operators that deviate from the wave equation only by the presence of weak dispersion, as are known to arise in models of quantum gravity, where a thoroughgoing understanding of QNMs is of special interest.
The modelling of dispersive gravitational wave propagation and its influence on the observable QNM spectrum will be essential if black hole spectroscopy is to be an effective probe into the domain of quantum gravity.

A further motivation for generalizing the non-relativistic method is to shed light on QNM spectral instabilities, and facilitate experimental tests of the recent ultraviolet universality conjecture, which posits that sufficiently high overtones converge to logarithmic Regge branches in the complex plane, in the high-frequency limit of potential perturbations \cite{Jaramillo2021, Boyanov2024}.
This effect is easily seen in numerical calculations of the P\"{o}schl-Teller spectrum, on account of its simplicity, but has yet to be experimentally confirmed.
Using the optical soliton, whose perturbations realize this potential, experimental tests become possible.
The numerical method presented above is essential for the modelling of these experiments, as one cannot realize an exact soliton in practice, and must always work with near-soliton potentials.
In addition, higher-order dispersive effects will also be present in any experiment, and these must be understood in order to interpret observations of QNM spectral migration with the soliton.
In particular, the influence of weak third-order dispersion acting on the perturbative probe field should be incorporated into the analysis, in order to provide the best test of the above conjecture.
This motivates the development of the non-relativistic method beyond the Schr\"{o}dinger equation, to include higher-order dispersive terms.

In view of the above reasons to generalize the non-relativistic method, we present a sketch of the more general method, which we will elaborate in future work.
Suppose we have a non-relativistic equation of the form
\begin{equation}
(\alpha(i\partial_t) + \beta(i\partial_r) + V)\phi = 0,\label{GeneralEquation}
\end{equation}
with $\alpha(z)$ and $\beta(z)$ finite polynomials in $z$, and $d$ the larger degree among the two polynomials.
In principle, we can apply a hyperboloidal coordinate transformation and a phase rotation of the fields, parameterised by the asymptotic velocities, $v_g$ and $v_p$.
Then, we introduce auxiliary fields to effect a $d$th-order reduction in time, defining
\begin{equation}
\phi_1 = \phi, \quad \phi_{k+1} = \partial_\tau\phi_{k},\label{ReductionInTime}
\end{equation}
with $1\leq k<d$.
Eq.~(\ref{ReductionInTime}) closely mirrors the treatment of resonator QNMs in optics \cite{lalanne2019}.
The equations of motion of these fields are trivial for all fields but $\phi_d$, whose equation of motion more closely resembles Eq.~(\ref{InitialEquationOfMotion}). If we use a Taylor series expansion of $\partial_\tau\phi_d$ around zero, we can write it in terms of spatial operators acting on the fields. The general form of the now $d$-dimensional operator $L$ is
\begin{equation}
L = 
\begin{pmatrix}
0 & 1 & 0 & \cdots & 0\\
0 & 0 & 1 & \cdots & 0\\
0 & 0 & 0 & \cdots & 0\\
\vdots & \vdots & \vdots & \ddots & 1\\
L_1 & L_2 & L_3 & \cdots & L_d\\\label{LMatrix}
\end{pmatrix},
\end{equation}
which we discretize as before. Then, we use the asymptotic dispersion relation of Eq.~(\ref{GeneralEquation}) to eliminate the asymptotic velocities, obtaining a vector equation for the QNM frequencies.
From Eq.~(\ref{LMatrix}), it can be shown that it is always possible to construct an $N$-dimensional matrix $M$ whose determinant is a finite polynomial for the QNM frequencies.
This may then be solved numerically and the frequencies $\Omega$ determined.
This generalization
is largely straight-forward. However, the divergences in space are multi-exponential with higher derivatives, leading to non-polynomial modes in the compactified coordinates.
This complicates the imposition of QNM boundary conditions, and further work is required to address this.
For example, approaches that augment the function space to include additional non-polynomial functions can be investigated.
Future work can investigate how this generalized method compares with other numerical schemes, as the connection to the relativistic method is less concrete in this case.

The method presented is primarily intended for the gravitational context and long-range potentials, but the authors note that extensions to optical cavities or plasmonic resonators may be possible.
Beyond QNMs, the non-relativistic method can be applied to spectra of non-selfadjoint operators, connecting with a larger research effort.
We believe an explicit formulation in this context is a promising research direction.
In addition, future works can develop the method, along the lines of \cite{Jaramillo2021}, in order to calculate the pseudospectra of non-relativistic operators.
It is our view that the relationship between perturbed QNM spectra and the pseudospectrum is best understood from a broader perspective, not limited to relativistic wave operators.
We expect that numerical methods will become increasingly important for addressing questions in the theory of QNMs, and anticipate that investigations into the QNMs of non-relativistic fields will provide new avenues to explore these questions.

The supporting data for this Letter are openly available
from \cite{data_source}.

This work was supported in part by the Science and Technology Facilities Council through the UKRI Quantum Technologies for Fundamental Physics Programme [Grant ST/T005866/1].
CB was supported by the UK Engineering and Physical Sciences Research Council [Grant EP/T518062/1].

We would like to express our thanks to Th\'eo Torres for providing us a useful overview at the outset of this research.

\onecolumngrid
\appendix
\newpage
\section{Appendix}
\setcounter{equation}{0}
\renewcommand{\theequation}{A\arabic{equation}}
\subsection{1. Boundedness of QNMs on the Spatial Boundaries}
The new coordinates of Eqs.~(\ref{coordtransformation}) and (\ref{coordfuncs}) are a simple modification of Bizo\'{n}-Mach coordinates, parametrized with the asymptotic velocity, $v_g$.
The explicit coordinate transformation may be written as
\begin{equation}
\tau = t - (v_g)^{-1}\log\cosh (r), \quad x = \tanh (r),
\end{equation}
If we consider the asymptotic behaviour of these coordinate functions as $r\to\pm\infty$, then we find that
\begin{equation}
\tau \to t \pm (v_g)^{-1}(r + \log(2)),
\end{equation}
In the original coordinates, this reads
\begin{equation}
\tau \to (v_g)^{-1}(v_g t \pm r) \pm (v_g)^{-1}\log(2),\label{consttau}
\end{equation}
from which we see that contours of constant $\tau$ correspond to trajectories propagating outwards with velocity, $v_g$.

The asymptotic form of the QNM solutions may be written
\begin{equation}
\phi \sim \exp(-\text{Im}(K) r + \text{Im}(\Omega) t)\exp(i\text{Re}(K)r - i\text{Re}(\Omega) t),\label{asympmodesplit}
\end{equation}
where we have separated out evolution of the amplitude and phase. For complex frequencies and wavenumbers, we generalize the group velocity, given by $\text{Im}(\Omega)/\text{Im}(K)$, to correspond to the transport of amplitude.

The boundedness of QNM solutions on the boundaries of the compactified space follows when we consider QNMs whose group velocities at the right and left boundaries are $v_g$ and $-v_g$, respectively.
The contours of constant $\tau$ define the compactified space, and it follows from Eqs.~(\ref{consttau}) and (\ref{asympmodesplit}) that the asymptotic evolution of amplitude along these contours vanishes for QNMs with the stated group velocities.
As a result, these QNMs take on definite amplitudes at the boundaries of the compactified space, $x = \pm 1$, and are bounded everywhere on the space. 

\subsection{2. Relationship between $\Omega$ and the Asymptotic Velocities}
The asymptotic dispersion relation associated with Eq.~(\ref{schrEqn}) is $\Omega = -K^2$.
The frequency and wavenumber are both complex quantities in the case of QNMs, and we begin by separating real from imaginary parts, to obtain
\begin{equation}
\text{Re}(\Omega) = (\text{Im}(K)^2 - \text{Re}(K)^2), \quad 
\text{Im}(\Omega) = -2\text{Re}(K)\text{Im}(K).
\end{equation}
Inverting these to obtain the real and imaginary parts of the wavenumber, we arrive at
\begin{equation}
\text{Re}(K) = \pm\sqrt{\frac{1}{2}(|\Omega| - \text{Re}(\Omega))}, \quad
\text{Im}(K) = \pm\sqrt{\frac{1}{2}(|\Omega| + \text{Re}(\Omega))},
\end{equation}
where the sign choices are correlated. As discussed above, Eq.~(\ref{asympmodesplit}) shows that the asymptotic group velocity is the ratio of the imaginary parts of the frequency and wavenumber. Likewise, it shows that the asymptotic phase velocity is the ratio of their real parts. Identifying the parameters $v_g$ and $v_p$ with the group and phase velocities at the right spatial boundary, $x = 1$, we obtain
\begin{equation}
v_p = \frac{\text{Re}(\Omega)}{\text{Re}(K_+)} = -\frac{\sqrt{2}\text{Re}(\Omega)}{\sqrt{|\Omega| - \text{Re}(\Omega)}}, \quad
v_g = \frac{\text{Im}(\Omega)}{\text{Im}(K_+)} = -\frac{\sqrt{2}\text{Im}(\Omega)}{\sqrt{|\Omega| + \text{Re}(\Omega)}}.
\end{equation}
These give rise to a pair of equations,
\begin{equation}
2\text{Re}(\Omega)^2 - v_p^2 |\Omega| + v_p^2\text{Re}(\Omega) = 0, \quad 2\text{Im}(\Omega)^2 - v_g^2|\Omega| - v_g^2 \text{Re}(\Omega) = 0,
\end{equation}
which relate the real and imaginary parts of $\Omega$ to the real parameters, $v_g$ and $v_p$. Requiring QNM solutions to decay in time and have outgoing group velocities at the boundaries, we can solve for $\Omega$ and obtain Eq.~(\ref{FrequencyInTermsOfSpeeds}).

\subsection{3. Equation of Motion for the $\hat{\psi}$ Field}
\noindent The auxilliary field $\hat{\psi}$ obeys Eq.~(\ref{InitialEquationOfMotion}), with
\begin{align}
J_1 &= v_g^2\Big[V + \Delta(1-x^2) - \Delta^2 x^2 + 2(1+\Delta)x(1-x^2)\partial_x - (1-x^2)^2\partial_x^2\Big], \nonumber\\
J_2 &= v_g\Big[iv_g + 1 - x^2 - 2\Delta x^2 + 2x(1 - x^2)\partial_x \Big].
\end{align}
In order to obtain Eq.~(\ref{EquationOfMotion}) of the main text, we first split $J_1$ and $J_2$ by whether $x^2$ is a factor, writing
\begin{align}
x^2\partial_\tau\hat{\psi} = x^2(L_{1}^a\hat{\phi} + L_{2}^a\hat{\psi}) + J_{1}^b\hat{\phi} + J_{2}^b\hat{\psi},\label{SplitEquation}
\end{align}
in which we define the new operators as
\begin{align}
L_1^a &= v_g^2\Big[\tilde{V} - (1 + \Delta)\Delta - 2(1+\Delta)x\partial_x + (2-x^2)\partial_x^2\Big],
&L_2^a = v_g\Big[{-}1 - 2\Delta - 2x\partial_x \Big], \nonumber\\
J_1^b &= v_g^2\Big[V_0 + xV_1 + \Delta + 2(1+\Delta)x\partial_x - \partial_x^2\Big],
&J_2^b = v_g\Big[iv_g + 1 + 2x\partial_x \Big],\hspace{8pt}
\end{align}
with $\tilde{V} = (V-V_0-xV_1)/x^2$ given by
\begin{equation}
\tilde{V} = \sum_{k=0}^{\infty}\frac{x^{k}}{(k+2)!}V_{k+2}.
\end{equation}
By writing $\partial_\tau\hat{\psi} =\partial_\tau\hat{\psi}_{a} + \partial_\tau\hat{\psi}_{b}$, we can partition the field so that
\begin{align}
\partial_\tau \hat{\psi}_{a} &= L_{1}^a\hat{\phi} + L_{2}^a\hat{\psi}, \label{nondivQNMeqn}\\
x^2\partial_\tau \hat{\psi}_{b} &= J_{1}^b\hat{\phi} + J_{2}^b\hat{\psi},
\label{divQNMeqn}
\end{align}
satisfying Eq.~(\ref{SplitEquation}) by construction. We note that Eq.~(\ref{nondivQNMeqn}) is already in the desired form of Eq.~(\ref{EquationOfMotion}), whereas Eq.~(\ref{divQNMeqn}) reproduces the form of Eq.~(\ref{InitialEquationOfMotion}). Nevertheless, we can express $\partial_\tau\hat{\psi}_b$ in the form
\begin{equation}
\partial_\tau\hat{\psi}_b = L_1^b\hat{\phi} + L_2^b\hat{\psi},
\end{equation}
using a Taylor series expansion about $x = 0$. In particular, by repeated differentiation of Eq.~(\ref{divQNMeqn}), we can obtain all spatial derivatives of $\partial_\tau\hat{\psi}_b$ at $x = 0$. In this way, we obtain the operators $L_1^b$ and $L_2^b$ via
\begin{equation}
L_1^b\hat{\phi} = \sum_{k = 0}^{\infty}\frac{x^k}{(k+2)!} \int^{+1}_{-1}\!\!\text{d}x\ \!\delta(x)(\partial_x^{n+2}J_1^b\hat{\phi}), \quad
L_2^b\hat{\psi} = \sum_{k = 0}^{\infty}\frac{x^k}{(k+2)!} \int^{+1}_{-1}\!\!\text{d}x\ \!\delta(x)(\partial_x^{n+2}J_2^b\hat{\psi}),
\end{equation}
where integration with the Dirac delta function evaluates the derivatives at $x = 0$. Explicitly, we find
\begin{align}
L_1^b &=
v_g^2 \sum_{k = 0}^{\infty}\frac{x^k}{(k+2)!} \int^{+1}_{-1}\!\!\text{d}x\ \!\delta(x)\big[(k+2) V_1\partial_x^{k+1} + (V_0 + 4 + 2k + \Delta(5+2k))\partial_x^{k+2} - \partial_x^{k+4}\big]\ \!\bigcdot\nonumber\\
L_2^b &=
v_g \sum_{k = 0}^{\infty}\frac{x^k}{(k+2)!} \int^{+1}_{-1}\!\!\text{d}x\ \!\delta(x)\big[(5 + 2k + iv_g)\partial_x^{k+2}\big]\ \!\bigcdot
\end{align}
where $\bigcdot$ indicates that the operand forms part of the integrand. The end result is
\begin{equation}
L_1 = L_1^a + L_1^b, \quad L_2 = L_2^a + L_2^b,
\end{equation}
which provides the operators appearing in Eq.~(\ref{EquationOfMotion}) of the main text.

\subsection{4. Determinant of the Matrix $L^N - \Omega \text{Id}$}
\noindent The determinant of the matrix $L^N - \Omega \text{Id}$ may be written as
\begin{equation}
\det(L^N - \Omega \text{Id}) = \det
\begin{pmatrix}
-\Omega \text{Id} & i\text{Id} \\
iL_1^N & iL_2^N - \Omega \text{Id}
\end{pmatrix}
= \det(\Omega^2 \text{Id} - i\Omega L_2^N - L_1^N),
\end{equation}
where we have used that $\Omega \text{Id}$ is invertible.

The QNM spectrum is given when this determinant vanishes, meaning we can scale $\Omega^2 \text{Id} - i\Omega L_2^N - L_1^N$ without changing our result. First, we substitute for the discretized operators $L_1^N$ and $L_2^N$, making use of the asymptotic dispersion relation to rewrite $v_g$ and $v_p$ in terms of the real and imaginary parts of $\Omega$. Then, after an overall scaling of the matrix, we arrive at the condition $\det(M) = 0$, with
\begin{align}
M &= \tilde{V}^N
- \sqrt{\Omega}(\sqrt{\Omega}+1)I^N
+ 2(\sqrt{\Omega}+1) X^N D^N
+ \big(2-(X^N)^2\big)(D^N)^2 \nonumber\\
&\quad\quad +\sum_{k = 0}^{N-2}\frac{(X^N)^k}{(k+2)!} \delta^N \Big[(k+2) V_1(D^N)^{k+1} + [V_0 + (\sqrt{\Omega} + 2(k + 2))(\sqrt{\Omega}+1)](D^N)^{k+2} - (D^N)^{k+4}\Big],\label{Mexpression}
\end{align}
where $X^N$ is a diagonal matrix of Chebyshev extremal points and $D^N$ is the corresponding Chebyshev differentiation matrix. $\tilde{V}^N$ is a diagonal matrix of $\tilde{V}$ evaluated at those same points, and $\delta^N$ is the unique matrix representation of integration with the Dirac delta function, obtained by considering its effect on polynomial interpolants.

We see from the form of Eq.~(\ref{Mexpression}) that $\det(M)$ is a polynomial of degree $2N$ in $\sqrt{\Omega}$, with real coefficients, meaning that the solutions $\sqrt{\Omega}$ come in complex conjugate pairs.
\end{document}